\documentclass[12pt]{article}
\title{\Large \bf  \Large  On the world cognizability}
\author{B.L.Ioffe \\
A.I.Alikhanov Institute of Theoretical and Experimental Physics (ITEP), \\
B.Cheremushkinskaya 25, 117218, Moscow, Russia}
\begin{document}
\date{}
\maketitle

\newcommand{\be}{\begin{equation}}
\newcommand{\ee}{\end{equation}}

\def\la{\mathrel{\mathpalette\fun <}}
\def\ga{\mathrel{\mathpalette\fun >}}
\def\fun#1#2{\lower3.6pt\vbox{\baselineskip0pt\lineskip.9pt
\ialign{$\mathsurround=0pt#1\hfil##\hfil$\crcr#2\crcr\sim\crcr}}}
\newcommand{\bosigma}{\mbox{\boldmath $\sigma$}}

\vspace{5mm}

\begin{abstract}

The paper gives a few examples of the phenomena that will never be
understood by the mankind. The first example is the physics at the
scales of $\sim 10^{-33}$ cm where the gravitation interaction
becomes strong, sales at the very beginning of the Big Bang. It
argues that the mankind will never establish the laws that controls
the events at these scales. Further, it is supposed that the time
dependence of the entropy, which determine the direction of the time
arrow, originates at the same time scale and, thus, the nature of
the time arrow will be never established either. Finally, I
conjecture that the brain cells are controlled by quantum computer
with the very large (or even infinite) number of degenerate states.
An external observation destroys this degeneracy, leading to
impossibility to understand the mechanism of the conscience.

\end{abstract}

\vspace{5mm}

The philosophers are often discussing  the problem of the world
cognizability. The various philosophical concepts give different
answers to the question.

In this short paper I try to demonstrate, that there are a number
of incognizable phenomena i.e. the ones that their  nature and
mechanism will never be understood by the mankind. I will consider
two physical examples and one examples, based on the hypothetical
mechanism of the brain operation.

I start from elementary particle physics.  For long time the dream
of theoretical physicists  was the creation of the united theory,
which would describe all interactions -- strong, electromagnetic,
weak and gravitational. About 30 years ago, when the string theory
of gauge field was formulated, many physicists started  to hope that
this discovery opened  the way to  such theory. Unfortunately, as it
became clear recently, the number of string theories is incredibly
high (the estimates of their number give 10$^{500}$ or even
10$^{1000}$), see e.g. [1], so it is impossible to choose among
them the correct one. Even worse, after 30 years of work, the string
theory could not make a any physical predictions thus failing the
main test for any scientific theory. All this shows that  a purely
theoretical effort cannot succeed in creation of the united theory.

On the experimental side, currently the available energies extend up
to energy of Large Hadron Collider (LHC) $\sim $14 TeV, which
correspond to distances large than $10^{-18}$ cm. The study the
scales $\ga 10^{-18}$ cm confirm the Standard Model, which units
weak and  electromagnetic interactions (in the form of electroweak
theory) and include4s the theory of strong interaction  -- QCD.
According to the preliminary data the Higgs boson with the mass 125
GeV was found. With the discovery of the Higgs boson the electroweak
theory become complete and selfconsistent.

The gravitational interaction has a unique position among
fundamental interactions.  Its characteristic  scales, the
distances, at which the gravitational  interaction becomes strong
are of order $10^{-33}$ cm. The region between $10^{-18}$ cm and
$10^{-33}$ cm will never be investigated  experimentally because the
construction of an accelerator with the energies, corresponding to
the distances $10^{-33}$ cm, i.e. with energies about $10^{16}$ TeV,
is impossible: the Earth resources are not sufficient for it.
(Indeed, the power consumed by LHC is of order 100 MWt. The
increasing of accelerator energy up to $10^{16}$ TeV would require
the power $\sim 10^{17} $ MWt=$10^{11}$ TWt, whereas the power of
world electric power stations in now (2--3) TWt.) The domain of
distances from $10^{-18}$ cm up to $10^{-33}$ cm is not a barren
desert as is indicated by the existence of neutrino mass and
oscillations. Another indications is the abundance of the dark
matter in the Universe. These facts are not accounted for by the
Standard model. Thus, one cannot hope that in the energy domain
10--$10^{16}$ TeV these is nothing besides already known
interactions. Since this domain will be never studied experimentally
(expect at its lower border), one comes to the conclusion that
physics of smallest scales in incognizable.

The other example refer to problem of entropy. According to the
second law of thermodynamics all processes in the system lead to
the increase of the entropy, or, in the case of total equilibrium,
do not charge it. The increase of the entropy with time is not
connected with the structure of the Lagrangian, since the
Lagrangian of all interactions (expect of weak interaction) are
invariant under time inversion and the role of weak interaction is
negligible in macroscopic processes.

Let us consider the processes occurring in the Universe and go
backwords in time to the moment of the Big Bang. S. Weinberg [2] had
shown that known physical laws(which  are symmetric under exchange
$t \to -t$) are sufficient to describe the evolution of the Universe
starting from the time $10^{-4}$ sec after Big Bang. At the moment
$t=10^{-4}$ sec the entropy of  the observed part of the Universe is
only one order of magnitude smaller  than its modern
value\footnote{The author is indebted to S.I.Blinikov for the
estimation.} It is expected, that at the initial stage of Big Bang,
the entropy of the world was very small. Such a possibility was
formulated by Penrose [3] and considered also in Landau and Lifshitz
``Statistical Physics'' [4]. Therefore, the huge increase of entropy
that sets the arrow of time, took place in a small time interval
after Big Bang, when gravitational effects were strong and the
quantum effects were  significant, i.e. when the space-time
quantization played an important  role. This problem was discussed
many times (see, e.g. [3]), but the result was unsatisfactory: no
consistent explanation of the problem was found. If the increase of
the entropy, determining the time arrow took place in the domain
$\sim 10^{-33}$, then in accordance with what is said above, its
origin will remain unclear to us.

Notice that incognizability is not a new concept in science, it is
well established in mathematics where it is known that in some
theories it is principally  impossible to find if a given statement
is true of false. There is the famous G\"odel theorem, proved in
1930 [5], which  states: for any system of axioms of arithmetics one
can find a statement, about which it is impossible to say if it is
true of false (see also [6]).

As a final example, let us consider now the work of the cells of the
brain. It is clear, that there is some  mechanism of control in
these cells. E. Liberman [7] conjectured that the control of cells
(or their associations) is due to a molecular computer. The
similarity of the work the cells of the brain and the computers is
discussed in details by Penrose  [3]. I want
 to argue that this molecular computer should be a quantum
 computer. The main argument for this is that the quantum  computer is
 using not only the wave function amplitude, but also its phase,
 i.e.  the quantum computer allows to get much more information
 than its classical analogue and as  a consequence has much more
 extended mechanism for management. The similar property has the
 brain. The other argument for the quantum computer is the
 phenomenon  of the will freedom. The human (or the animal)
 operates by taking decisions, the choice among alternative
 decisions is very quick  and, probably, do not require any
 additional energy. This resembles the wave function in quantum
 mechanics that collapses to one of ground states under
 observation. Clearly, in the cell computer the number of such
 ground states must be very large. Such quantum computer might
 resemble that the non-abelian quantum gauge field theory: in which
 an infinitely large number of degenerate levels are distinguished
  by special topological (or winding) numbers  (see, e.g. [8]).
  The transitions between such degenerate levels are tunnel
  transitions and various mixing of such levels is possible [9].

  All this makes me believe that the brain    relies on the quantum
  computation which involves a lot of generate levels and that it
  is the existence of these levels that determines the conscience.
  Of course, it is a very strong hypothesis. One corollary of it
  is the incognizability  of the conscience. Indeed, any external
  intervention introduces the energy, destroys the quantum phases
  of levels and, as consequence destroys the conscience.

  The conclusion is that there are incognizable in Nature.

  This work was supported by RFBR grant 12-02-00351.


\vspace{1cm}


\begin{thebibliography}{99}

\bibitem{1} G. Brumfiel, Nature {\bf 443} 491 (2006).
\bibitem{2} S. Weinberg, {\it The first three minutes}, FLAMINGO,
Fontana, 1976.
\bibitem{3} R. Penrose, {\it The Emperors New Mind}, Oxford Univ. Press,
1999.
\bibitem{4} L. Landau, E. Lifshitz, {\it Statistical Physics}, part 1,
Pergamon Press, 1970.
\bibitem{5} K. G\"odel, Monatshefte  f\"ur Mathematic and Physik,
{\bf 38}, 173 (1931).
\bibitem{6} V. A. Uspensky, {\it Teorema G\"odelya o nepolnote},
M. Fizmalit, 1982; V. A. Uspensky, {\it Teorema G\"odelya o
nepolnote i 4 dorogy, vedushie k nei}, VII Letnya shkola
``Sovremennaya matenatika", JINR 2007.
\bibitem{7} E. Liberman, Biophysica {\bf 17} (1972) 932.
\bibitem{8} B. L. Ioffe, V. S. Fadin anf L. N. Lipatov, {\it Quantum
Chromodynamics, Perturbative and Nonperturbative Aspects}, Cambr.
Univ. Press, 2011.
\bibitem{9} B.V. Geshkenbein and B.L. Ioffe, Nucl. Phys. {\bf B166}
(1980) 340.




\end{thebibliography}
\end{document}